\documentclass{elsart}    

\usepackage{graphicx}          

\begin{document}

\begin{frontmatter}

\title{A parametric study of the lensing properties of dodecagonal photonic quasicrystals}


\thanks[footnoteinfo]{This paper was  presented at the PECS VII Conference.
Corresponding author: E. Di Gennaro, Tel. +39-081-7682661, Fax
+39-081-2391821.}

\author[Napoli]{E. Di Gennaro}\ead{emiliano@na.infn.it},
\author[Benevento]{D. Morello},
\author[Napoli]{C. Miletto},
\author[Napoli]{S. Savo},
\author[Napoli]{A. Andreone},    
\author[Benevento]{G. Castaldi},
\author[Benevento]{V. Galdi},
\author[Benevento]{V. Pierro}

\address[Napoli]{CNISM and Department of Physics, University of Naples
``Federico II'', \\ Piazzale Tecchio 80, I-80125, Napoli, ITALY}

\address[Benevento]{Waves Group, Department of Engineering, University of Sannio,\\Corso
Garibaldi 107, I-82100 Benevento, ITALY}


\begin{abstract}                          
We present a study of the lensing properties of two-dimensional
(2-D) photonic quasicrystal (PQC) slabs made of dielectric cylinders
arranged according to a 12-fold-symmetric square-triangle aperiodic
tiling. Our full-wave numerical analysis confirms the results
recently emerged in the technical literature and, in particular, the
possibility of achieving focusing effects within several frequency
regions. However, contrary to the original interpretation, such
focusing effects turn out to be critically associated to local
symmetry points in the PQC slab, and strongly dependent on its
thickness and termination. Nevertheless, our study reveals the
presence of some peculiar properties, like the ability to focus the
light even for slabs with a reduced lateral width, or beaming
effects, which render PQC slabs potentially interesting and worth of
deeper investigation.
\end{abstract}

\end{frontmatter}

\section{Introduction}
Photonic crystals (PCs), featuring a periodically modulated
refractive index, represent an active research area since nearly
two decades \cite {Yablonovitch,John}. The strong interest toward
these structures is motivated by their capability of inhibiting the
spontaneous emission of light, which potentially renders them the
building-block components of an all-optical circuitry \cite{Shinya}.

In PCs, multiple wave scattering at frequencies near the Bragg
condition prevents propagation in certain directions, producing a
{\em bandgap}. Anomalous (even {\em negative}) refraction and
subwavelength focusing (``superlensing'') properties for PCs have
been theoretically predicted (see, e.g.,
\cite{Notomi,Soukoulis,Lu,Luo}) and experimentally verified (cf.
\cite{Parimi,Cubukcu}). The theoretical foundations of the (analytic
and numerical) modeling, prediction, and physical understanding of
the above phenomena are based on well-established tools and concepts
developed for the study of wave dynamics in {\em periodic}
structures, such as Bloch theorem, unit cell, Brillouin zone,
equifrequency surfaces (EFS), etc.

With specific reference to lensing applications, two different
approaches have been presented to obtain subwavelength resolution
using a PC slab. The first one assumes that, under suitable
conditions (single beam propagation, negative group velocity in the
specific frequency band, and circular EFS), a PC can behave like a
homogeneous material with a negative refractive index $n=-1$. These
features are likely found near a frequency band edge and in the case
of high dielectric contrast \cite{Notomi}. Furthermore, modifying
the slab terminations, the impedance mismatch with the surrounding
medium can be minimized \cite{Decoopman} and/or surface waves can be
excited allowing (partial) reconstruction of evanescent waves
\cite{Moussa2}. Such a slab is not a perfect lens \cite{Pendry}, in
the sense that the image has a finite resolution, since not all
evanescent components can be restored. Nevertheless, the focus
position follows the simple ray-optical construction as for a flat
lens with $n=-1$ \cite{Lu,Wang1}. The second approach relaxes some
of the previous requirements (negative group velocity and isotropic
refraction properties) and achieves ``all-angle negative
refraction'' (AANR) without an effective negative index, provided
that the EFS are all convex and larger than the frequency contour
pertaining to the surrounding medium \cite{Luo}. This condition is
still sufficient for a PC slab to focus with subwavelength
resolution, if the termination is designed so as to sustain surface
modes. It is worth noticing that, in this case, the focus position
does not follow the ray-optical construction \cite{Li} and is
\emph{restricted} \cite{Wang1}.

Recently, there has been an increasing interest toward the study of
{\em aperiodically ordered} PCs. Such structures are inspired by the
``quasicrystals'' in solid-state physics \cite{Shechtman,Levine} and
the related theory of ``aperiodic tilings'' \cite{Senechal}, and are
accordingly referred to as ``photonic quasicrystals'' (PQCs). Like
their solid-state physics counterparts, PQC structures can exhibit
long-range (e.g., quasiperiodic) order and weak (e.g., statistical
and/or local) rotational symmetries which are not bounded by the
crystallographic restriction and can thus be considerably {\em
higher} than those achievable in periodic PCs \cite{Zoorob}.
 Accordingly, it has been proposed to
utilize PQC structures in order to maintain the periodic-like
scattering of light, while introducing additional geometric degrees
of freedom potentially useful for performance control and
optimization. However, it must be said that the lack of periodicity
renders the study of PQCs very complex and
computationally-demanding. Although some concepts developed for
periodic PCs can be used for the analysis of the properties of PQCs
\cite{Kali2,Kali}, a rigorous extension of a Bloch-type theorem (and
associated tools and concepts) does not exist.

Recent studies of the EM properties of 8-fold
\cite{Liu}, 10-fold \cite{Kali2,DellaVilla} and 12-fold
\cite{Zoorob} PQCs have shown that a gap in the density of states,
accompanied by significant dips in the transmission spectra, can be
achieved. PQCs also allow, in principle, for a higher degree of flexibility and
tunability for defect-type \cite{Ozbay,Cheng} and intrinsic
\cite{DellaVilla2} localized modes. Moreover, examples of
subwavelength focusing properties have been presented for 2-D PQCs
exhibiting various (8-fold, 10-fold, and 12-fold) rotational
symmetries \cite{Feng,Zhang2}. Similar results were also obtained
in the case of acoustic waves \cite{Zhang3}.

Although the above studies on the focusing properties of PQC slabs
reveal some appealing properties, such as polarization-insensitive
and non-near-field imaging, we believe that the original
phenomenology interpretation provided in \cite{Feng}, within the
framework of ``effective negative refractive-index and evanescent
wave amplification,'' is questionable and deserves a deeper
investigation.
 In this paper, we present some representative results
from a comprehensive parametric study of the focusing properties of
a 12-fold symmetric PQC slabs (as in \cite{Feng}), based on
full-wave numerical simulations (Fourier-Bessel multipolar
expansion). In particular, with reference to the line-source imaging
scenario, we investigate the critical role of the global vs. local
symmetry of the lattice (which is merely glossed over in
\cite{Zhang2}), by varying the PQC slab thickness, lateral width,
and termination, as well as the source position in both parallel and
orthogonal directions with respect to the slab surface.

\section{Samples and Methods}
\begin{figure}
\begin{center}
\includegraphics [width=0.65\textwidth]{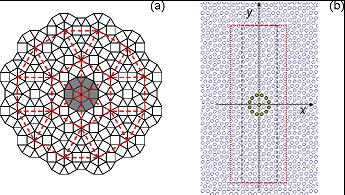}
\caption{Dodecagonal PQC geometry. (a): Illustration of the Stampfli
inflation rule. The gray-shaded dodecagon in the center represents
the parent tiling from which a big parent (red dashed lines) is
generated by applying a scaling factor $\nu=\sqrt{3}+2$. (b): Two
examples of PQC slabs of different thickness (black-dotted and
red-dashed rectangles) extracted from the tiling. The
green-full-dots dodecagon in the center corresponds to the parent
tiling in (a).} \label{geom}
\end{center}
\end{figure}

As in \cite{Feng}, the 2-D PQC of interest is made of
infinitely-long dielectric cylinders with relative permittivity
$\varepsilon_r$ = 8.6 and radius $r =0.3a$ placed (in vacuum) at the
vertices of a 12-fold-symmetric aperiodic tiling with lattice
constant (tile side-length) $a$, generated by the so-called Stampfli
inflation rules \cite{Oxborrow} illustrated in Fig. \ref{geom}(a).
The (recursive) generation algorithm is based on a parent tiling
represented by the central gray-shaded dodecagon. Scaling up this
structure by an inflation factor $\nu = \sqrt{3}+2$, one obtains a
big parent (red dashed dodecagon in Fig. \ref{geom}(a)). Copies of
the original tiling are subsequently placed at each vertex of the
big parent, and the process is iterated up to the desired tiling
extension. From a suitably-sized tiling, we extract several PQC
slabs of different sizes, with a thickness $t$ (along the $x$-axis)
ranging between $\sim 7a$ and $\sim 11a$ (see the two rectangles in
Fig. \ref{geom}(b), corresponding to the two examples presented in
\cite{Feng}) and a lateral width $h$ (along the $y$-axis) varying
between $\sim 11a$ and $\sim 30a$.
Our numerical simulations are carried out via a 2-D full-wave method
based on a multipolar expansion of the fields around (and inside)
each cylinder \cite{Felbacq}. This method is closely related to the
Korringa-Kohn-Rostocker method used in solid-state physics
\cite{Kohn} (which is itself derived from a pioneering work by Lord
Rayleigh in electrostatics \cite{Rayleigh}), and has also been
recently applied to the study of the local density of states in
finite-size PCs \cite{Asatryan,Asatryan2} and PQCs
\cite{DellaVilla}. In the present investigation, only E-type
polarization is considered, i.e. the electric field is assumed to be
parallel to the cylinders.

\section{Representative Results}
As a preliminary step, we studied the focusing properties of the two
PQC slabs in Fig. \ref{geom}(b), for a fixed line-source position,
and varying the normalized frequency $a/\lambda$ (with $\lambda$
being the vacuum wavelength) within a range between the first and
second bandgaps ($0.354\le a/\lambda\le 0.533$). We found several
frequency regions (including those in \cite{Feng}) where a clear
focus was visible. We note that this feature represents a first
remarkable difference with the PC case, where the focusing regions
tend to be more rare and well separated. In PQCs, likely due to
their inherent self-similar nature \cite{Kali}, focusing effects
seem to occur more densely in the spectrum. In all simulations
below, attention is focused on the normalized frequency
$a/\lambda=0.394$, corresponding to the value for which
a refractive behavior emulating a homogeneous medium with $n=-1$
was estimated in \cite{Feng}.

\subsection{Effects of slab thickness}
In order to gain some insight into the focusing properties, we then
considered PQC slabs of different thickness. The line-source was
always placed at a distance $d_x=t/2$ from the vacuum-slab
interface, and in correspondence of the center of the parent tiling
($y=0$). From a simple ray-optical construction (for flat lenses
with $n = -1$), the focus is accordingly expected to be
located in a symmetric position with respect to the slab.
\begin{figure}
\begin{center}
\includegraphics [width=0.65\textwidth]{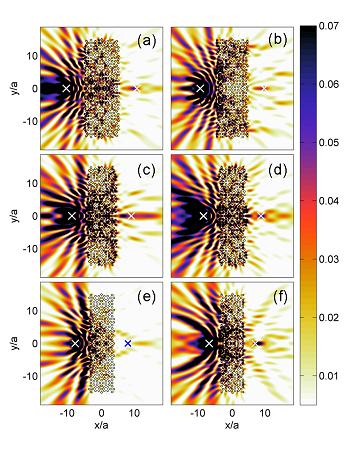}
\caption {Intensity field maps for various symmetric slab
configurations with different values of the thickness $t$. (a):
$t=11a$; (b): $t=9.8a$; (c): $t=9a$; (d): $t=8.8a$; (e): $t=8a$;
(f): $t=7a$. An electric line-source is placed at $y=0$ and a
distance $d_x=t/2$ from the vacuum-slab interface. The crosses mark
the source position and the ray-optical ($n=-1$) prediction for the
focus location. The color-scale limits in the plots are chosen so as
to properly reveal details in the internal and external fields that
would be otherwise overwhelmed by the dominant features.}
\label{simm}
\end{center}
\end{figure}
The slab thickness was changed by removing cylinders in such a way
to preserve the inversion symmetry with respect to the $y$-axis,
thereby obtaining six different configurations, which include as
extremes the two samples shown in Fig. \ref{geom}(b). As a first
result, our simulations confirmed the results in \cite{Feng}, as
shown in the (intensity) field maps in Figs. \ref{simm}(a) and
\ref{simm}(f). A focus is clearly observable at a position very
close to the ray-optical prediction (marked by a cross symbol).
However, for intermediate thickness (see Figs.

\ref{simm}(b)--\ref{simm}(e)), the focusing properties become rather
questionable. Moreover, in all the simulations, the reflection
coefficient in the source plane is considerably nonzero, indicating
the presence of a significant impedance mismatch at the vacuum-slab
interface. Furthermore, the field distributions inside the slab turn
out to be significantly different, suggesting a strong dependence
from the whole slab configuration, including its termination. In
this connection, we note that it is not possible to change the PQC
lens thickness without changing at least one termination. In Figs.
\ref{simm}(c) and \ref{simm}(d), a feature that resembles a
``beaming effect'' can be observed, together with a strong field
intensity along the vacuum-slab interface on the image side. In
periodic PC slabs, such a phenomenon has been related to the
presence of localized photon states on the slab surface \cite{Luo},
stemming from an \emph{overall resonance} in the structure.
\begin{figure}
\begin{center}
\includegraphics [width=0.65\textwidth]{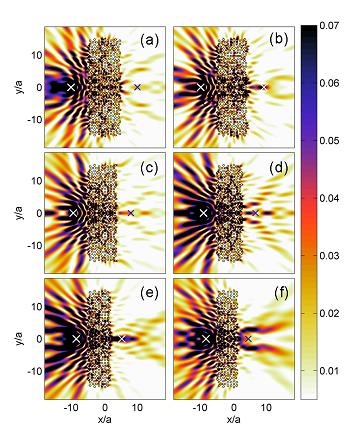}
\caption {As in Fig. \ref{simm}, but for asymmetric configurations.
(a): $t=10.3a$; (b): $t=9.8a$; (c): $t=9a$; (d): $t= 8a$; (e):
$t=7a$; (f): $t=6.6a$.} \label{asimm}
\end{center}
\end{figure}

In Fig. \ref{asimm}, the field maps pertaining to other six
different PQC slabs of various thickness are displayed. In these
examples, the slab thickness was changed by removing dielectric
cylinders only from the image side, thereby destroying the inversion
symmetry with respect to the $y$-axis. Also in this case, the field
distribution on the image side appears to depend critically on the
structure geometry. Although a clear single focus cannot be
identified, several spots (Figs. \ref{asimm}(a) and \ref{asimm}(d)),
or beaming effects (Figs. \ref{asimm}(b), \ref{asimm}(e), and
\ref{asimm}(f)) are visible, indicating that the inversion symmetry
does play a key role in the focus formation.

\subsection{Effects of source position}
\begin{figure}
\begin{center}
\includegraphics [width=0.65\textwidth]{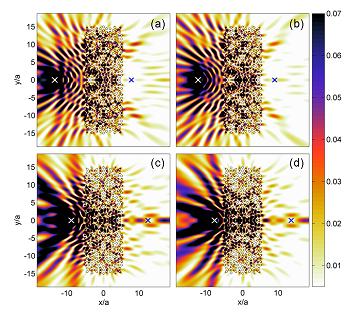}
\caption {As in Fig. \ref{simm}(a) (i.e., thicker slab, cf. red
dashed rectangle in Fig. \ref{geom}(b)), but for various source
distances $d_x$ from the vacuum-slab interface. (a): $d_x=8.3a$;
(b): $d_x=7.1a$; (c): $d_x=3.6a$; (d): $d_x=2.5a$. The source
$y$-position is maintained at $y=0$. } \label{ortog}
\end{center}
\end{figure}

We then fixed the slab geometry to that of Fig. \ref{simm}(a)
(thicker slab), that was found to exhibit focusing properties, and
changed the source position. In case of a flat lens with  $n = -1$,
simple ray-optics predicts the source-image distance to remain
constant and equal to twice the lens thickness \cite{Lu}. Results
for periodic PCs showing an almost isotropic refractive response
(see, e.g., \cite{Wang1}) agree fairly well with this prediction.
First, we kept the source in a symmetric position along the
transverse direction ($y = 0$), varying the distance $d_x$ from the
slab from $0.75a$ up to $10.5a$. In Fig. \ref{ortog}, some
representative field maps are shown. A variety of effects is
observed, ranging from the absence of focus (Fig. \ref{ortog}(a)) to
multiple spots (Figs. \ref{ortog}(c) and \ref{ortog}(d)), but always
in the near-field region, raising further concerns about the
interpretation in \cite{Feng}. Similar results \cite{Li} were also
observed for PCs, in the case of an anisotropic regime with positive
or negative refraction properties, where the near-field focusing was
found to be dominated by a self-collimation effect.

\begin{figure}
\begin{center}
\includegraphics [width=0.65\textwidth]{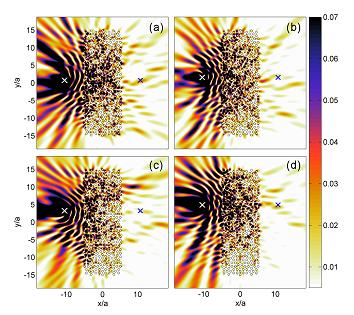}
\caption {As in Fig. \ref{ortog}, but for various source
displacements $d_y$ along the $y$-axis. (a) $d_y=0.75a$; (b):
$d_y=1.5a$; (c): $d_y=3a$; (d): $d_y=6a$. The source distance from
the vacuum-slab interface is maintained at $d_x=5.5a$ (i.e., half
the slab thickness). } \label{parall}
\end{center}
\end{figure}

\begin{figure}
\begin{center}
\includegraphics [width=0.65\textwidth]{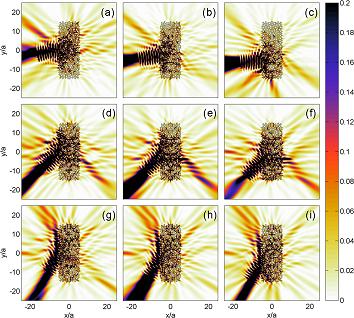}
\caption {As in Fig. \ref{simm}(a), but for Gaussian-beam incidence
at various lateral positions $y_i$ at the slab interface, and
various incidence angles $\theta_i$ (with respect to the $x$-axis).
The Gaussian beam minimum spot-size is $\sim 2.2a$ and the incidence
point is located at the Rayleigh distance. (a): $y_i=0$,
$\theta_i=10^o$; (b): $y_i=-2.44a$, $\theta_i=10^o$; (c):
$y_i=-4.88a$, $\theta_i=10^o$; (d): $y_i=0$, $\theta_i=45^o$; (e):
$y_i=-2.44a$, $\theta_i=45^o$; (f): $y_i=-4.88a$, $\theta_i=45^o$;
(g): $y_i=0$, $\theta_i=60^o$; (h): $y_i=-2.44a$, $\theta_i=60^o$;
(i): $y_i=-4.88a$, $\theta_i=60^o$.} \label{slab}
\end{center}
\end{figure}

Next, for a fixed source-slab distance $d_x=t/2$, we moved the
source parallel to the vacuum-slab interface. Representative results
are shown in Fig. \ref{parall}. In PCs, for source displacements
that preserve the distance from the slab and the structure
periodicity, the focus remains unaffected, and its position follows
the source location. Conversely, in our PQC case, when the $x$-axis
symmetry is broken, the focus undergoes a rapid deterioration, and
is completely destroyed for a source displacement of $d_y = 6a$ (see
Fig. \ref{parall}(d)). These findings highlight the importance
of keeping the source on the same axis of the central symmetry point
in the 12-fold tiling (highlighted in Fig. \ref{geom}(b)), as also observed in \cite{Zhang2}.
For the thicker slab configuration in Fig. \ref{geom}(b), we also
performed some simulations involving a collimated Gaussian beam
impinging on three different points of the PQC slab interface with
three different incidence angles. Results shown in Fig. \ref{slab},
displaying complex multi-beam features in the transmitted field that
are strongly dependent on the incidence point, highlight the absence
of a clear-cut refractive behavior, and explain the strong
dependence of the focusing effects on the source distance from the
slab (cf. Fig. \ref{ortog}) and lateral displacement (cf. Fig.
\ref{parall}).

\subsection{Effects of slab lateral width}
\begin{figure}
\begin{center}
\includegraphics [width=0.65\textwidth]{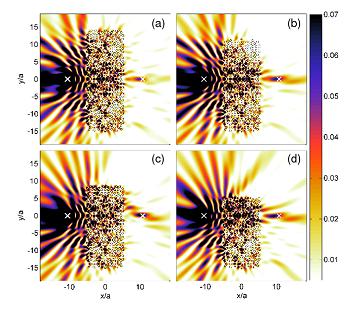}
\caption {As in Fig. \ref{simm}(a), but for various asymmetric slab
configurations with different lateral width $h$. (a): $h=29.4a$;
(b): $h=26.7a$; (c): $h=23.9a$; (d): $h=20.6a$.} \label{ridux}
\end{center}
\end{figure}

\begin{figure}
\begin{center}
\includegraphics [width=0.65\textwidth]{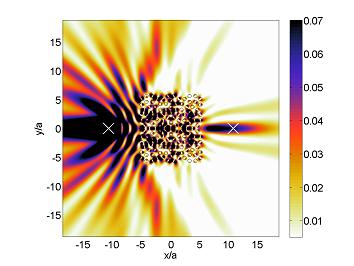}
\caption {As in Fig. \ref{ridux}, but reducing the lateral width to
$\sim 11a$ symmetrically with respect to the $y$-axis.}
\label{slab_foc}
\end{center}
\end{figure}

To better illustrate the role of the local symmetry point, we also
varied the slab lateral width asymmetrically (i.e., removing
cylinders only from one side of the $y$-axis) while keeping the
source in the central position ($d_x=t/2$ and $y=0$). Results are
shown in Fig. \ref{ridux}. In this case, the focus turns out to be
rather stable, in spite of the $x$-symmetry breaking. These results
clearly reveal the key role played in the focus formation by {\em
short-range} interactions involving a neighborhood of the parent
tiling (highlighted via green full dots in Fig. \ref{geom}(b)).
This can be also observed considering a configuration where the
slab lateral width is symmetrically reduced to $\sim 11a$ with
respect to the $y$ axis, as shown in Fig. \ref{slab_foc}. Again, in
spite of the considerably reduced lateral width, the focusing
properties associated to the symmetry local point located at
$(x=0,y=0)$ appear to be rather robust, as if their occurrence was
restricted to a limited range of incidence angles. This renders
these PQC structures particularly appealing for the development of
compact optical systems.

\section{Conclusions}
To sum up, a comprehensive parametric numerical study of the lensing
properties of quasiperiodic dodecagonal PQC slabs has been carried
out. Results confirm that these structures possess focusing
properties in several frequency regions, even if critically
associated to local symmetry points (and neighboring regions)
existing in the tiling lattice. Indeed, our full-wave numerical
simulations show that varying the slab thickness or the source
position, the field distributions inside the slab and in the image
plane change dramatically. The original interpretation in
\cite{Feng}, within the framework of ``effective negative
refractive-index and evanescent wave amplification,'' does not
appear adequate to describe the underlying phenomenologies, which
entail complex near-field scattering effects and short-range
interactions. Nevertheless, this particular PQC shows some
peculiarities in the EM response that deserve further investigation.
In particular, applications to lenses with small aperture or
antennas with highly directional beaming could be envisaged.

Also of interest it is the parametric study of the refraction and
focusing properties of other classes of PQCs based on periodic
(e.g., Archimedean \cite{David}) and aperiodic (e.g., Penrose
\cite{Kali2,DellaVilla}, octagonal \cite{Liu}, circular \cite{Xiao})
tilings characterized by diverse degrees of (local and/or
statistical) rotational symmetry.

\begin{ack}                               
This work has been funded by the Italian Ministry of Education and
Scientific Research (MIUR) under the PRIN-2006 grant on ``Study and
realization of metamaterials for electronics and TLC applications.''
The authors wish to thank the referees for their useful comments and
suggestions, as well as  Mr. V. De Luise and Mr. A. Micco for their technical support.  
\end{ack}


\begin{thebibliography}{10}
\expandafter\ifx\csname url\endcsname\relax
  \def\url#1{\texttt{#1}}\fi
\expandafter\ifx\csname urlprefix\endcsname\relax\def\urlprefix{URL
}\fi

\bibitem{Yablonovitch}
E.~Yablonovitch, Phys.\ Rev.\ Lett 58 (1987) 2059.

\bibitem{John}
S.~John, Phys.\ Rev.\ Lett. 58 (1987) 2486.

\bibitem{Shinya}
A.~Shinya, S.~Mitsugi, T.~Tanabe, M.~Notomi, I.~Yokohama, H.~Takara,
  S.~Kawanishi, Optics Express 14 (2006) 1230.

\bibitem{Notomi}
M.~Notomi, Phys. Rev. B 62 (2000) 10696.

\bibitem{Soukoulis}
S.~Foteinopoulou, C.~M. Soukoulis, Phys. Rev. B 72 (2005) 165112.

\bibitem{Lu}
W.~Lu, S.Sridhar, Optics Express 13 (2005) 10673.

\bibitem{Luo}
C.~Luo, S.~Johnson, J.~Joannopoulos, J.~Pendry, Phys.\ Rev.\ B 65
(2002)
  201104R.

\bibitem{Parimi}
P.~Parimi, W.~Lu, P.~Vodo, S.Sridhar, Nature 426 (2003) 404.

\bibitem{Cubukcu}
E.~Cubukcu, K.~Aydin, E.~Ozbay, S.~Foteinopoulou, C.~M. Soukoulis,
Nature 423
  (2003) 604.

\bibitem{Decoopman}
T.~Decoopman, G.~Tayeb, S.~Enoch, D.~Maystre, B.~Gralak, Phys. Rev.
Lett. 97
  (2006) 073905.

\bibitem{Moussa2}
R.~Moussa, T.~Koschny, C.~Soukoulis, Phys. Rev. B 74 (2006) 115111.

\bibitem{Pendry}
J.~B. Pendry, Phys. Rev. B 85 (2000) 3966.

\bibitem{Wang1}
X.~Wang, Z.~Ren, K.~Kempa, Optics Express 12 (2004) 2919.

\bibitem{Li}
Z.~Li, L.~Lin, Phys. Rev. B 68 (2003) 245110.

\bibitem{Shechtman}
D.~Shechtman, I.~Blech, D.~Gratias, J.~W. Cahn, Phys. Rev. Lett. 53
(1984)
  1951.

\bibitem{Levine}
D.~Levine, P.~J. Steinhardt, Phys. Rev. Lett. 53 (1984) 2477.

\bibitem{Senechal}
M.~Senechal, Quasicrystals and {G}eometry, Cambridge University
Press,
  Cambridge (UK), 1995.

\bibitem{Zoorob}
M.~Zoorob, M.~Charleton, G.~Parker, J.~Baumberg, M.~Netti, Nature
404 (2000)
  740.

\bibitem{Kali2}
M.~Kaliteevski, S.~Brand, R.~Abram, T.~Krauss, R.~D.~L. Rue,
P.~Millar,
  Nanotechnology 11 (2000) 274.

\bibitem{Kali}
M.~Kaliteevski, S.~Brand, R.~Abram, T.~Krauss, R.~D.~L. Rue,
P.~Millar, J.\
  Mod.\ Opt. 47 (2000) 1771.

\bibitem{Liu}
Y.~S. Chan, C.~T. Chan, Z.~Liu, Phys.\ Rev.\ Lett. 80 (1998) 956.

\bibitem{DellaVilla}
A.~D. Villa, S.~Enoch, G.~Tayeb, V.~Pierro, V.~Galdi, F.~Capolino,
Phys.\ Rev.\
  Lett. 94 (2005) 183903.

\bibitem{Ozbay}
M.~Bayndir, E.~Cubukcu, I.~Bulu, E.~Ozbay, Phys.\ Rev.\ B 63 (2001)
161104(R).

\bibitem{Cheng}
S.~Cheng, L.-M. Li, C.~Chan, Z.~Zhang, Phys.\ Rev.\ B 59 (1999)
4091.

\bibitem{DellaVilla2}
A.~D. Villa, S.~Enoch, G.~Tayeb, F.~Capolino, V.~Pierro, V.~Galdi,
Optics
  Express 14 (2006) 10021.

\bibitem{Feng}
Z.~Feng, X.~Zhang, Y.~Wang, Z.~Li, B.~Cheng, D.~Z. Zhang, Phys.\
Rev.\ Lett. 94
  (2005) 247402.

\bibitem{Zhang2}
X.~Zhang, Z.~Li, B.~Cheng, D.~Zhang, Optics Express 15 (2007) 1292.

\bibitem{Zhang3}
X.~Zhang, Phys.\ Rev.\ B 75 (2007) 024209.

\bibitem{Oxborrow}
M.~Oxborrow, C.~Henley, Phys.\ Rev.\ B 48 (2003) 6966.

\bibitem{Felbacq}
D.~Felbacq, G.~Tayeb, D.~Maystre, J. Opt. Soc. Am. A 11 (1994) 2526.

\bibitem{Kohn}
W.~Kohn, N.~Rostoker, Phys. Rev. 94 (1954) 1111.

\bibitem{Rayleigh}
L.~Rayleigh, Philos. Mag 34 (1892) 481.

\bibitem{Asatryan}
A.~Asatryan, K.~Busch, R.~McPhedran, L.~Botten, M.~de~Sterke,
N.~Nicorovici,
  Phys. Rev. E 63 (2001) 046612.

\bibitem{Asatryan2}
A.~Asatryan, S.~Fabre, K.~Busch, R.~McPhedran, L.~Botten,
M.~de~Sterke,
  N.~Nicorovici, Optics Express 8 (2001) 191.

\bibitem{David}
S.~David, A.~Chelnokov, J.~Lourtioz, IEEE J.\ of Quantum Elect. 37
(2001) 1427.

\bibitem{Xiao}
S.~Xiao, M.~Qiu, Photonics and Nanostructures –- Fundamentals and
Applications
  3 (2005) 134.

\end{thebibliography}
\end{document}